# Synthetic topological device for advancing elastic energy harvesting

Jiamin Guo[1], Zhongming Gu[1]*, Lei Fan[2], Jie Liu[1], Yafeng Chen[1], Zhongqing Su[2]†, & Jie Zhu[1]‡

[1]*Institute of Acoustics, School of Physics Science and Engineering, Tongji University, Shanghai, 200092, China*

[2]*Department of Mechanical Engineering, The Hong Kong Polytechnic University, Hung Hom, Kowloon, Hong Kong SAR, China*

## Abstract

High-efficiency energy harvesting of ultrasonic elastic waves are crucial for powering electric gadgets in many emerging technologies such as wearable devices, wireless sensing, and biomedical implants. Although topological phononic metamaterials have recently been demonstrated as a promising paradigm for confining and guiding elastic waves through robust bound states, achieving ultrahigh-$Q$ topological resonance with enhanced energy conversion efficiency remains a challenge. In this work, we propose a synthetic-dimensional higher-order topological insulator by engineering the flexural bands of elastic metamaterials, featuring highly localized topological hinge states in the bulk bands. This topological hinge mode stems from the nonzero combination of the bulk polarization and the Chern number in the synthetic-dimensional band structure, thus giving rise to a strong elastic-to-electric energy conversion at the corner of the phononic plate. Through numerical simulations and experimental validations, straightforward evidence of the localized modes with robust protection and consequent abilities in activating the light-emitting diodes (LEDs) array have been demonstrated. Our findings open a new avenue for topological-physics-enabled ultrasonic devices and present promising prospects for applications in weak-signal detection and self-powered sensors.

# I. INTRODUCTION

Squeezing energy from the environment is a long pursuit in research community and plays a vital role in cutting-edge technologies, spanning from robotics, microelectronic systems to biomedical engineering. Among various energy forms, elastic waves offer distinct advantages for on-chip and embedded energy harvesting, attributed to their efficient propagation through diverse media[1-4], biocompatibility and safe operation within human tissues[5,6], and robust performance against environmental perturbations including thermal fluctuations and electromagnetic interference[7-9]. In particular, high-frequency ultrasonic elastic waves are increasingly valued for their short wavelengths and pinpoint directionality, enabling precise spatial concentration and high energy density[1,10-13], which are fundamental requirements for miniaturized systems. Previous efforts have largely focused on low-frequency vibrations due to their lower energy losses[14,15], harvesting energy from high-frequency ultrasonic waves encounter many challenges such as vibration dissipation[12,16], wave scattering[17-19], and inefficient electromechanical coupling[5,6,20], significantly limiting the performance of compact devices operating at micro-nano scale. Overcoming these limitations is crucial for enabling efficient energy conversion in next-generation technologies.

In addition to increasing the electromechanical coupling coefficient of piezoelectric materials, achieving strong spatial confinement of elastic-wave energy is a critical strategy for enhanced energy harvesting[21,22]. Many mechanisms, such as structural focusing and impedance matching[23-25], have been utilized to confine elastic waves yet suffer from significant scattering losses at high frequencies. The advent of elastic topological metamaterials (ETMs), an extension of concepts from electronic systems to classical wave systems, provides significant insights into the novel wave-matter interactions. By sharing similar band structures, the intricate topological notions can be verified in elastic platforms and inspiring a series of intriguing wave behaviors[26,27], such as nonreciprocal propagation, bandgap-induced localization, and mode conversion suppression. Among them, symmetry-protected edge[28-33] and corner states[34-39] can guide mechanical waves with minimal backscattering, thereby suppressing defect-induced losses and enhancing robustness against disorders. However, the confinement performance of topological boundary states is inherently limited by the bandgap size, resulting in inadequate energy density for efficient harvesting. Another choice is embedding the topological boundary states in the continuum of bulk bands without leakage, which remains unexplored for high-frequency elastic waves. Addressing these challenges is therefore essential to unlock the full potential of ETMs for ultrasonic energy harvesting.

In this work, we propose a synthetic three-dimensional (3D) topological insulator engineered to harness its higher-order topological bound states in the continuum (HOTBICs) for high-frequency ultrasonic elastic energy harvesting. The designed elastic structure emulates an extra spatial dimension by introducing a controlled phase shift into the intracell couplings. Periodic modulation of the hopping along this synthetic axis generates an effective gauge field, endowing the bulk band structure with a higher-dimensional topological invariant[40-42]. Notably, although the design is intrinsically 3D in the real-synthetic space, its physical implementation is simplified to a two-dimensional (2D) structure, with a projection of topological hinge states in the bulk bands. Numerical optimization of the modulation parameters ensures maximal localization of the elastic wave in the bound modes, without the hybridization with the bulk modes. Subsequent elastic-wave experiments further verify the ultrahigh-$Q$ performance of HOTBICs compared to conventional topological bound states, and their effectiveness for enhanced energy harvesting. Our results also show that the energy conversion efficiency of this elastic topological device remains stable under lattice defects, validating the topological protection mechanism. Such a robust platform for highly localized energy harvesting is well suited to be integrated into compact devices in the fields of wearable health monitoring and passive structural sensing.

## II. RESULTS

### A. Tight-binding model and numerical results

Here, Fig. 1 conceptually illustrates the theoretical model of proposed ETM and its mechanism for efficient ultrasonic energy harvesting. In Fig. 1a, the unit cell of the tight-binding lattice consists of $2 \times 3$ sites: coupling along the $y$-direction varies with the phase parameter $\phi$, while coupling along the $x$-direction alternates periodically between strengths $\tau_1$ and $\tau_2$. By sweeping $\phi$ over a full $2\pi$ cycle, a 3D Chern insulator is realized in the ($k_x$, $k_y$, $\phi$) real-synthetic space, with its topological character defined by the Chern number associated with $\phi$. Thus, it is possible to construct higher-order topological states, associated with strong energy confinement, in a low-dimension space. Meanwhile, by judiciously adjusting the structural parameters, the flexural bands of elastic crystal can be mapped to this lattice mode, whose strong surface localization can generate large elastic strains. As shown in Fig. 1b, these localized vibrational waves are then converted into electrical energy by a piezoelectric harvester with high efficiency. This fascinating behavior arises since the synthetic topological bound states can be tuned to exist in the bulk bands, thereby inheriting the characteristics of both topological states and the bound states in the continuum (BICs)[43-45]. Consequently, they achieve stronger spatial confinement of energy within the continuous spectrum, compared to the resonance based on single mechanism. Such a combined feature allows the ETM to accumulate the strain energy even in the presence of defects or disorders, yielding strong elastic-to-electric conversion.

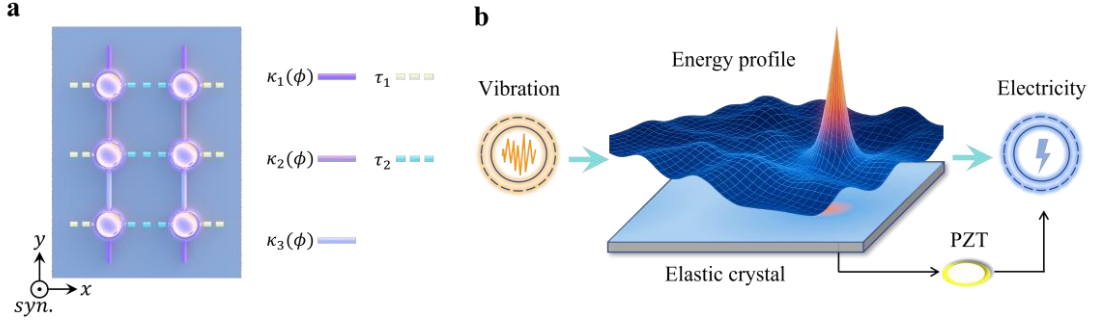

**Fig. 1: Schematic diagram of the elastic topological device. a** A basic unit in the lattice model of the designed ETM. In the $y$ direction, hopping terms $\kappa_1$, $\kappa_2$, and $\kappa_3$ are modulated by the synthetic momentum $\phi$, while the $x$ direction exhibits staggered couplings. **b** The lattice model in (**a**) can be constructed practically by an elastic crystal, whose energy profile is illustrated to be highly localized. The piezoelectric patch (PZT) converts elastic energy to electrical energy.

Considering the open boundary conditions in both the $x$ and $y$ directions, the lattice model can be described by the following Hamiltonian:

$$H = \sum_{i,j}[t_{(i,j),(i+1,j)}\hat{c}^\dagger_{i,j}\hat{c}_{i+1,j} + t_{(i,j),(i,j+1)}\hat{c}^\dagger_{i,j}\hat{c}_{i,j+1} + \mathrm{H.\,c.}], i,j \in (1,N), \qquad (1)$$

where $N$ is the number of sites along each axis ($x$ and $y$), and $\hat{c}^\dagger_{i,j}$ ($\hat{c}_{i,j}$) represents the creation (annihilation) operator for a spinless particle at the site $(i,j)$. The hopping amplitude along $x$ alternates according to

$$t_{(i,j),(i+1,j)} = (-1)^i \frac{\tau_2-\tau_1}{2} + \frac{\tau_1+\tau_2}{2}, \qquad (2)$$

while the coupling along $y$ is modulated as

$$t_{(i,j),(i,j+1)} = \kappa_0 + \delta\kappa\cos(2\pi bj + \phi), \qquad (3)$$

with modulation frequency $b = 1/3$, unperturbed hopping amplitude $\kappa_0$, modulation strength $\delta\kappa$, and phase factor $\phi$. The parameter $\phi$ acts as a synthetic dimension since $\phi$ corresponds to a periodic modulation orthogonal to the real-space lattice coordinates in parameter space[41,45,46], providing an independent axis. Thus, this 2D tight-binding model effectively describes a 3D Chern insulator in the combined real-synthetic space.

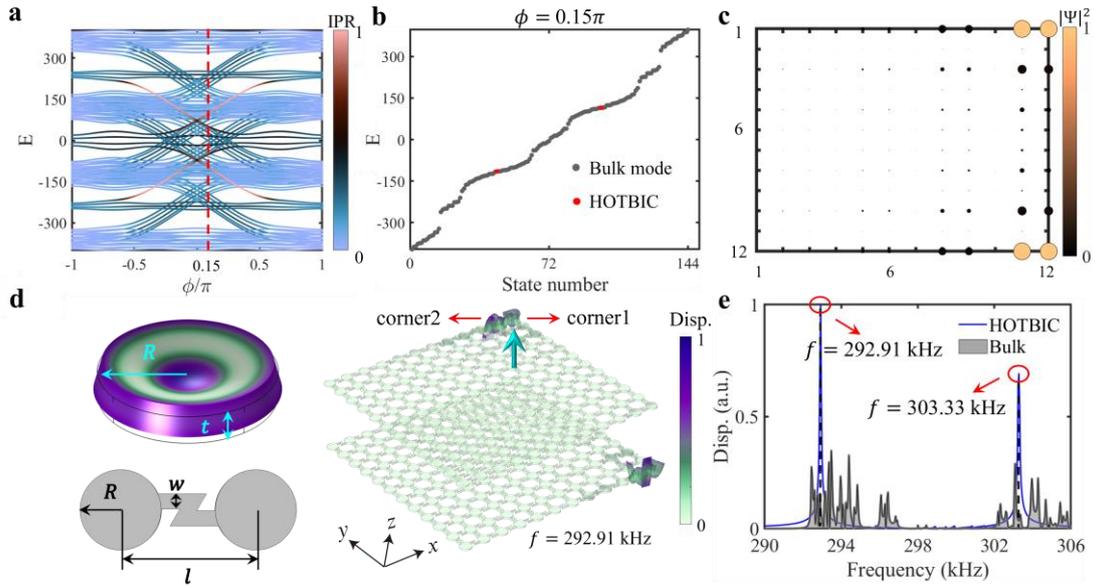

**Fig. 2: The analysis of tight-binding model and simulations of the ETM. a** Eigen spectra of the topological Chern insulator with finite size in both $x$ and $y$ directions ($12 \times 12$ sites). The color bar indicates the IPR of the corresponding eigenstates. **b** The eigen spectrum for the tight-binding model when phase parameter $\phi = 0.15\pi$. **c** The spatial distribution of the square of the wave-function amplitude for the HOTBICs in (**b**). **d** Left: the second-order flexural eigenmode of a disk and two disks connected with a Z-shaped beam. Right: simulated displacement fields of the excited synthetic topological hinge state in the continuum within the ETM. The arrow marks the source position. **e** The blue curve represents the simulated displacement response when the excitation and probe placed at corner1 and corner2 in (**d**) respectively. The gray shaded region corresponds to the response under the bulk excitation.

For a finite $12 \times 12$ ($N = 12$) lattice system, with $\kappa_0 = 120$, $\delta\kappa = 96$, $\tau_1 = 30$ and $\tau_2 = 120$, Fig. 2a presents the eigen-spectrum as a function of the synthetic parameter $\phi$. When $\phi = 0.15\pi$, two topological bound states appear within the continuum, displaying significantly higher inverse participation ratio (IPR) values compared to other modes. Here, IPR is employed as a metric for energy localization with the definition

$$\text{IPR} = \frac{\sum_i |\psi_i|^4}{(\sum_i |\psi_i|^2)^2}, \tag{4}$$

where $\psi_i$ denotes the mode amplitude at the $i$-th lattice site, so that modes with stronger spatial confinement is related to larger IPR values. Furthermore, Fig. 2b displays the eigen-spectrum along the red dashed line at $\phi = 0.15\pi$ in Fig. 2a, showing the existence of two twofold degenerate HOTBICs in the continuous bulk spectrum. Because of the chiral symmetry of the Hamiltonian in Eq. (1), the energy spectrum is strictly symmetric about zero energy, endowing the two nonzero-energy HOTBICS with identical energy distributions. As shown in Fig. 2c, the calculated spatial distribution of one of the HOTBICs reveals strong corner localization, which is consistent with the IPR analysis. Moreover, these modes exhibit strong robustness against disorder, as further discussed in the Supplemental Note 1 [47].

Based on the above analysis, a flexural-wave metacrystal then is proposed to realize the tight-binding model on an elastic platform, with finite-element simulations (COMSOL Multiphysics) confirming its behavior. The structure comprises identical circular disks (radius $R = 2.1$ mm, thickness $t = 0.7$ mm) interconnected by slender Z-shaped beams (width $w$, center-to-center spacing $l$), as depicted in Fig. 2d. The second-order flexural mode of each disk is selected as the on-site orbital and resonates around 297.43 kHz, whose vibration profile is radially symmetric with maximum amplitude occurring at the center and the perimeter. The coupling strengths between the adjacent disks can be tuned by changing the structural parameters $w$ and $l$.

In simulations, the band structure and eigenmodes of our designed ETM are shown to be equivalent to those of the tight-binding model (Supplemental Note 4 [47]). The right side of Fig. 2d presents the displacement distributions of the first synthetic topological hinge state at 292.91 kHz under corner excitation, exhibiting strong corner spatial localization, as predicted in Fig. 2c. This mode corresponds to the left sharp peak in the simulated response spectrum at corner2 under excitation of corner1, as shown in Fig. 2e. The second HOTBIC, associated with the right peak at 303.33 kHz, also demonstrates strong spatial confinement in Supplemental Fig. S6 [47]. In addition, the resonance peaks of the two HOTBICs fall within the gray-shaded region representing the bulk modes, verifying their property of being located in the bulk band. These results collectively indicate that the synthetic HOTBICs are highly confined in both spatial and spectral domains, highlighting their great potential in energy harvesting.

### B. Experimental validation

To conduct experimental validation, an ETM sample is laser-cut with overall dimensions of 82.81 mm × 81.62 mm × 0.7 mm, as shown in Fig. 3a. We capture the out-of-plane vibrational response spectrum at the center of the disk located at corner2 using the laser vibrometer, with excitation applied at corner1, as presented in Fig. 3b. It can be clearly observed that, there are two distinct peaks at $f_1 = 284.26$ kHz and $f_2 = 296.91$ kHz, indicating the existence of two topological corner states (Fig. 3d). Furthermore, the robustness of the ETM against impurities is also investigated by attaching several aluminum pillars with dimensions 5 mm × 5 mm × 2 mm on the ETM, set up as Supplemental Fig. S2 [47]. The results in Fig. 3c show that the response profiles with and without impurities tend to be similar, suggesting the strong robustness of the proposed ETM. During the experiments, a translational stage is used to scan the vibrational profiles of the entire lattice close to the peak frequencies. Figure 3d displays the measured displacement fields of three modes. From left to right, the first two panels correspond to the two corner-localized HOTBICs excited at frequency $f_1$ and $f_2$, respectively, via corner-driven excitation. The third panel shows a bulk mode measured at $f_1$ under bulk excitation, with the bulk mode at $f_2$ exhibiting similar distributions. The coexistence of corner-localized HOTBICs and bulk modes at the identical frequencies ($f_1$ and $f_2$) demonstrates that the former are bound states embedded within the continuous bulk spectrum. These observations align well with the obtained numerical results, validating the ETM's capability to faithfully emulate the targeted tight-binding Hamiltonian.

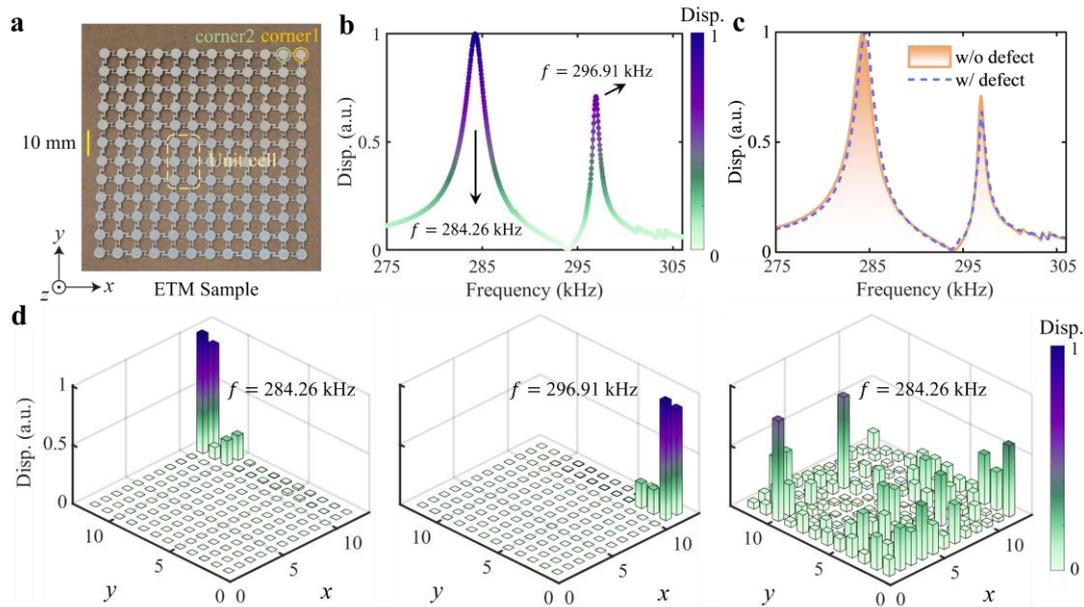

**Fig. 3: Experimental results. a** A photograph of the ETM sample consisting of a $12 \times 12$ array of disks. **b** The out-of-plane displacement spectrum measured at the center of corner2 in (**a**), under the excitation at corner1. **c** The measured displacement of the ETM with (purple solid line) and without (orange shadow) defects. **d** Measured out-of-plane displacement fields for the two HOTBICs and one bulk states. The height and color of each bar represent the normalized magnitude of the displacement.

### C. Ultrasonic energy harvesting and self-powered via the ETM

To convert the ultrasonic energy transmitted into electrical energy by the ETM, a piezoelectric harvester is affixed onto the response disk of the ETM, with a pure aluminum plate of identical dimensions serving as a reference (specific experimental details are provided in the Supplemental Note 6 [47]). Subsequently, we measured the output voltages as a function of load resistances at the frequency of 284.5 kHz and 294.5 kHz, which corresponds to the two HOTBICs in Fig. 2b, as plotted in Fig. 4a and Fig. 4b, respectively. It is observed that the output voltages increase with the load resistance and finally saturate at a certain threshold, holding the characteristics of a real voltage source. At the frequency of 284.5 kHz (case1), the saturation voltages probed at the ETM and bare plate are 0.424 V and 0.181 V, respectively, while at the frequency of 294.5 kHz (case2), they are 0.431 V and 0.182 V, respectively. Moreover, the output powers, calculated based on the Ohm's law, are presented in Fig. 4c and Fig. 4d. The maximum output powers harvested at the ETM and the bare plate under the case1 are 41.18 mW and 0.45 mW, while they are 30.73 mW and 0.22 mW under the case2, respectively. The maximum output powers harvested at the ETM in case1 and case2 are 91.5 and 139.7 times higher than those harvested at the bare plate, respectively, demonstrating the superior energy harvesting capacity of the HOTBICs. We also demonstrate the robustness of the energy harvesting performance against impurities in the Supplemental Note 2 [47]. Overall, these measured results conclusively demonstrate that the ETM's synthetic topologically induced localization mode not only amplifies energy conversion efficiency by over two orders of magnitude compared to a bare plate, but also maintains robust performance in the presence of structural impurities.

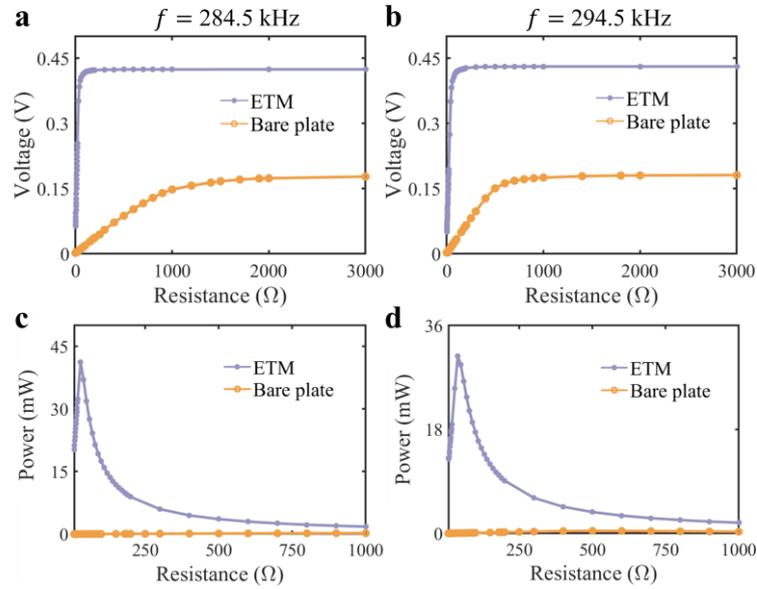

**Fig. 4: Ultrasonic energy harvesting. a, b** The output voltages versus load resistances for the ETM and bare plate with same size at (**a**) a frequency of 284.5 kHz and (**b**) a frequency of 294.5kHz, respectively. **c, d** The output powers versus load resistances for the cases that are same with (**a**) and (**b**), respectively.

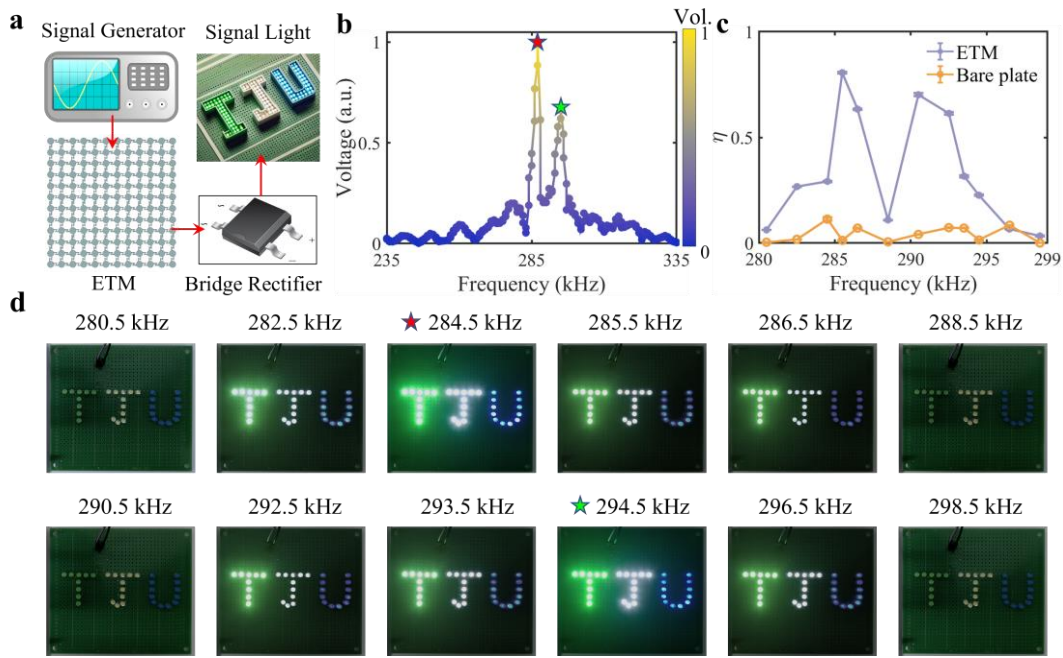

**Fig. 5: Self-powered and frequency-selective elastic topological device. a** Schematic diagram of the system. **b** The measured voltage response spectrum through the electric system under the corner excitation. **c** Energy conversion efficiency with error bars of the ETM and bare plate. **d** The experimental results that the luminescence level of LEDs changes as the excitation frequency.

Based on the highly efficient energy harvesting capability described above, a self-powered system capable of driving a large array of LEDs has been implemented. As depicted in Fig. 5a, a piezoelectric patch driven by a sweeping signal actuates on the ETM plate, and the resultant AC signal is converted into DC power through a rectifier bridge to supply 32 series-connected LED signal lights in the shape of "TJU". The harvested open-circuit voltage spectrum using the sine-sweep excitation signal is shown in Fig. 5b, coinciding well with the numerical results (Fig. 2e) and the experimental results via the laser vibrometer (Fig. 3b). As shown in Fig. 5c, the ETM-based system demonstrates two prominent efficiency peaks of 0.8 and 0.7 at frequencies of 284.5 kHz and 294.5 kHz, respectively. These efficiency values are calculated as the ratio of the output power (the electrical power supplied to the LED) to the input power provided by the sine-sweep excitation signal. The values significantly exceed those observed for the bare plate, aligning with the features in the output voltage spectrum. This enhancement further confirms the critical advantage of synthetic HOTBICs for harvesting elastic wave energy. To ensure the reliability and accuracy of the experimental findings, the experimental operation is carried out seven times to avoid the randomness of the results. Correspondingly, Fig. 5d illustrates the variations in LED brightness on the circuit board under an input signal within the 280.5-298.5 kHz frequency range. As the frequency of the signal generator increases, we can visually observe the changes in illuminance of LED array, with the highest brightness occurring at the frequencies of 284.5 kHz (red star) and 295.5 kHz (green star). The detailed evolutions of the brightness appear as a movie, which is provided in the Supplemental Note 7 [47]. Furthermore, long-term stability tests demonstrated that the harvested power and corresponding LED illumination remained consistently over repeated excitation cycles (more that 24 hours), thereby confirming the robustness and repeatability of our ETM-based energy harvesting mechanism.

## III. DISCUSSION

In this study, we present a synthetic-dimensional topological insulator that leverages higher-order topological hinge states in the continuum to enable efficient high-frequency ultrasonic elastic wave energy harvesting. The design enhances energy localization and reduces scattering losses and dissipation. Our results demonstrate superior performance in energy conversion compared to conventional systems, showcasing the robustness of synthetic topological hinge states in the continuum against structural defects. This is particularly significant in applications requiring compact and efficient systems, such as low-power sensors and wireless power transfer. The integration of piezoelectric energy conversion, i.e., the self-powered system that drives 32 LEDs, further underscores the practicality of our approach. Our experimental validation corroborates the numerical predictions, confirming the effectiveness of the topological design in real-world applications. While our current results focus on the 280-320 kHz regime, the underlying design principles are scalable to higher frequencies. Our work emphasizes the potential of topology in reshaping the landscape of high-frequency energy harvesting, opening new possibilities for sustainable and efficient power solutions across diverse fields.

# ACKNOWLEDGEMENTS

This work is supported by the National Natural Science Foundation of China (Grants No. 92263208 and No. 12304494), the National Key R&D Program of China (Grants No. 2022YFA1404400 and No. 2022YFA1404403), the Shanghai Science and Technology Committee (Grant No. 21JC1405600), the Research Grants Council of Hong Kong SAR (Grant No. AoE/P-502/20), the Fundamental Research Funds for the Central Universities.

[*]zhmgu@tongji.edu.cn

[†]zhongqing.su@polyu.edu.hk

[‡]jiezhu@tongji.edu.cn